\documentclass[12pt]{article}

\newcommand{\p}{\partial}

\newcommand{\aq}{\begin{eqnarray}}
\newcommand{\qa}{\end{eqnarray}}

\begin{document}

\begin{titlepage}

\begin{flushright}
math-ph/0506046 \\
UUPHY/05-06
\end{flushright}

\begin{center}

{\Large \bf{Classical symmetries of monopole by group theoretic methods}}

\vspace{10mm}

{\large{ Karmadeva Maharana } \\
\vskip 1.5cm
 {Physics Department, Utkal University, Bhubaneswar
  751 004, India}  \\

\sf karmadev@iopb.res.in  }

\vskip 3.5cm

\noindent

{{\bf{Abstract}}}
\end{center}


 { We use group theoretic methods to obtain the extended Lie point symmetries 
of the equations of motion for a 
charged particle in the field of a monopole. Cases with certain 
model magnetic fields and potentials are also studied.    
Our analysis gives the generators and Lie algebras  generating  
the inherent symmetries. The  equations of motion of a scalar
particle probing the near horizon structure of a black hole is also
treated likewise.  We have also found the generators of Krause's complete 
symmetry groups for some of the above examples. 
}

 
\end{titlepage}


\newpage
\section{Introduction}

In certain physical problems there may exist extra hidden symmetries
which are not apparent, unless searched for.
Some of the examples from classical considerations are,
a particle in a $1\over{r^2}$ potential in one dimension \cite{deA}, 
 the conserved Runge-Lenz vector
of the Kepler problem and the extra symmetries of
a charge moving in the field of a magnetic 
monopole\cite{{jackiw},{jackiw1}}, and the
 generators beyond the Poincare invariance that give rise to
conformal invariance in electrodynamics as well as in Yang-Mills theory. 
The existence of symmetries
help in classifying and obtaining energy levels and eigenstates
in quantum mechanical problems, generating new solutions
and also formulating conservation laws. As we know
manifestation of scale invariance in deep inelastic scattering
had deep significance in the development gauge theories. The invariance under
scale and conformal transformations also motivated the
construction of a simple 
classical model which leads to  conformal quantum mechanics\cite{deA}.
Recently there have been a revival of interest in this model. This is
due to the observation
in string theory dynamics that a particle near a black hole possesses
SO(2,1) symmetries as in conformal quantum mechanics.\cite{Sen}.

 The symmetries of charged particle - monopole system  and the 
conformal quantum mechanics were obtained from physical
reasonings and scale invariance. However, there
exists a general programme  to obtain
 the symmetries of the equations of motion of any such system by
 using the group theoretic methods of Lie. 
In this paper we use this method to find the Lie point symmetries 
of the monopole
system as well as some other physically motivated systems.

A knowledge of the symmetry group of a system of differential equations  
leads to several types of applications\cite{olver}.
For example, the symmetry group is helpful in finding solutions to the set of 
equations as well as
constructing new solutions to the systems from the known ones. 
There have been efforts to generalise and extend the Lie 
approach by
 considering non-standard symmetries, as well. This helps in getting 
a wider class of solutions. Non-classical symmetries result by 
the weakening of  invariance requirements
of the differential equation under the symmetry 
transformations\cite{nuccinc}. Another approach is 
to enlarge the 
space of independent variables by adding auxiliary variables 
and finding the
symmetries. Then the symmetries related to the original system are 
figured out
and the corresponding symmetries are 
called non-local\cite{{Ed},{Kras},{Leo}}. This way,
it also provides a method to classify different classes of 
solutions 
corresponding to different symmetries. One can use, on the 
other hand, 
symmetry groups to classify families of differential equations 
depending 
on arbitrary parameters or functions. As an extension of this 
idea, Krause 
has introduced the important idea of complete symmetry of a 
differential 
equation so as to expand the symmetry group such that  the 
manifold of 
solutions is an homogeneous space of the group and the 
group is specific 
to the system, i.e., no other system admits that symmetry group\cite{krause}. 
The complete 
symmetry group of the system is the group represented by 
the set of 
symmetries required to specify completely a system and
its point 
symmetries\cite{{nucci},{nuccilorenz},{nuccijacobi},{nuccimar}}.
 
  Further, just the enumeration of the symmetry generators sometimes 
provide much physical insight and quantitative physical results for 
which the full solutions are not required. The derivation of Kepler's 
third law of planetary motion
and Runge-Lenz vectors\cite{stephani}, calculation of energy levels
for hydrogen like atoms and generalized Kepler's problems, harmonic 
oscillators,  Morse 
potentials\cite{wybourne}, electron in a specific nonuniform
 magnetic field\cite{mah03}, 
being some  such examples. In these the energy eigenvalues are 
obtained through the method of spectrum generating
 algebras which gives the 
Casimir invariants directly without explicit recourse to the solutions.
Of course, the solutions are also obtained from the representation
theory.
For finding the continuous symmetries, Lie's method of 
group analysis seems to be the most powerful technique available.
For example,  Witten\cite{witten}
had considered an example of the equation of motion of a
 particle in three
dimensions constrained to move on the surface of a sphere in 
the presence
of a magnetic monopole. This is the classical analogue of the 
Wess-Zumino
model\cite{wess}.  The equations of motion of such a system in the
 presence of a magnetic monopole cannot be obtained from the usual
 Lagrangian
formulation unless one goes   to a higher dimension. Hence, the usual
method of finding the symmetries through  Noether's
 theorem would have
difficulties. So, to look for the continuous symmetries associated
 with such classical
systems one has to analyze directly the equations of motion\cite{mah01}.
Similar is the case for Korteweg-de Vries equation which is not
 amenable to a
direct Lagrangian formulation  when expressed as a  lowest order 
equation\cite{blumen}.
Another example is the Lorenz system of equations which have been dealt
in the papers by Sen and Tabor\cite{sen}, and Nucci\cite{nuccilorenz}. 
For the classical systems,
 this procedure of finding directly the symmetries from 
equation of motion
is, in some sense, more fundamental. This is because
in certain cases many different Lagrangians may  give
 rise to the same
equations of motion. The group analysis of the equations of 
motion gives
all the Lie point group symmetry generators. In the cases where 
a Lagrangian 
formulation is possible, the usual Noether symmetries are a subset of 
the above generators. This subset of generators acting on the
Lagrangian gives  zero\cite{stephani}.
However, besides these
there may be other generators obtained through group
analysis which have direct physical
significance, but not explicitly available from the consideration
of usual Noether symmetries alone\cite{prince}. The reproduction
of Kepler's third law in the planetary motion problem 
is such an example.
 The extension of this
idea to the notion of Lie dynamical symmetries contains
 similarly a subclass
known as Cartan symmetries. The Runge-Lenz vector can be 
obtained from
such considerations. These symmetries are, further, related
 to the Lie-B\"{a}cklund symmetries.

The application of these types of analysis to nonlocal cases have been
widely studied through B\"{a}cklund transformations and related
techniques in the context of integrable systems containing infinite
number of conservation laws\cite{ibragimov}. The Thirring model
 has been analyzed by
Morris \cite{morris}. The differential geometric forms developed earlier
 are used in the above analysis to obtain the 
prolongation structure\cite{estabrook}.
 Some other applications  of these ideas to important problems from 
physics is comprehensibly covered by Gaeta\cite{gaeta}.  

In this paper, as a first step towards a complete group analysis, we  
find the Lie point
symmetries of the coupled set of differential equations 
representing the motion of a charged particle in three dimensions:

(i)  in the presence of a magnetic field 
   proportional to the coordinate vector,

(ii)  in the presence of 
 a magnetic monopole stationed at the centre of a sphere, with the 
constraint that the particle moves on 
the surface of this sphere of unit radius,

(iii) in the presence of a $\frac{1}{r^2}$ potential

(iv)  in the presence of 
a magnetic monopole,
 
(v) in the presence of a dyon, 

(vi)  in a model magnetic field  of the form $ {\bf B }   
    = \{ 0, 0 , {-} {\frac{\cal B} {x^2}} \}$. Here 
  $ \cal B $ is a constant,

and

(vii) a particle in a  type of velocity dependent potential,

We also obtain the generators of the complete symmetry group 
of Krause   for
most of the above examples in this paper.

It should be noted that in the context of the symmetries of 
Wess-Zumino-Witten
 models, the  symmetries in the higher dimension play by far
the most important role and these have been fruitfully exploited  
 \cite{braaten}.  

Section 2 provides the outline of the method for group analysis of
the equations of motion. In section 3 we use this method to find the 
generators of the point symmetries for the above examples.
The symmetries of the equation of a scalar quantum particle near the
horizon of a massive blackhole is considered in section 4.
Next we follow the method of reduction of order introduced by 
Nucci\cite{{nucci},{nuccikepler2}}
to obtain the complete Krause symmetries for four of the above examples 
which is the content of section 5. Finally, Section 6 is devoted to
physical interpretations and conclusions.

\section{Symmetry Conditions}

 Typically we are interested in the coupled nonlinear set
of equations representing the equations of motion of a particle in
three dimensions. These are of the form
\begin{eqnarray}
  {{\ddot{x}}_{a}} = {\beta}{{\omega}_{a}} {({{x_{i}}},{\dot{x}}_{i},t )} 
   \label{eq:nfree}
\end{eqnarray}
where a dot represents derivative with respect to time, $ {a,i = 1,2,}$
and  $3$, and $ {\beta} $ is a constant involving mass,
coupling constant etc.. The expressions for the function ${\omega}_a $ will 
be given
explicitly for each example.
 
These set of equations can be analyzed by means of one parameter groups 
by infinitesimal transformations. We demand the equation to be invariant under
infinitesimal changes of the explicit variable ${t}$, as well as simultaneous
infinitesimal changes of the dependent functions ${x_{a}}$ in the following
way,

\begin{eqnarray}
  t \rightarrow {\tilde{t}} &=&  t + \epsilon {\tau} {(t, x_1, x_2, x_3)}
   + O {({\epsilon}^2)},
\nonumber\\
  {{x}_{a}} {\rightarrow} {\tilde{x}}_a &=& {{x}_{a}} + 
  {\epsilon}{{\eta}_{a} {(t, x_i)}}
   + {O {({\epsilon}^2)} } .\label{eq:zwei}
\end{eqnarray}

Under $t \rightarrow \tilde{t}$ and $x_a \rightarrow {\tilde{x}}$, the
equation changes to,
\begin{eqnarray}
  {\ddot {\tilde{x}}_a} &=& \beta {\tilde \omega}_a ({\tilde t }, {\tilde x}_i,
  {{\dot{\tilde x}_i}})
\end{eqnarray}

To illustrate the procedure consider the simple case in one space dimension.
We express the above equation in terms of ${ t} $ and ${q}$ by using the
transformation (\ref{eq:zwei}). Then the invariance condition implies 
that an expression
containing various partial derivatives of ${\tau}$ and ${\eta}$  is obtained
which equates to zero. For example, we get
\begin{eqnarray}
   { \frac{d{{\tilde x}}}{d{\tilde t}}} &=& { \frac{ dx + {\epsilon}
   {( {\frac{\p{\eta}}{\p{t}}} dt
   + \frac{\p{\eta}}{\p{x}}  dx )} }{ dt +
   {\epsilon}{(\frac{\p{\tau}}{\p{t}} dt
   + \frac{\p{\tau}}{\p{x}} dx )} }}   +  O ({\epsilon}^2) \label{eq:dxdt1}
\end{eqnarray}
and now relate the left hand side with $\frac{dx}{dt} $ by using binomial
theorem for the denominator to obtain 
\begin{eqnarray}
  \frac{d{\tilde{x}}}{d{\tilde t}}  &=&  \frac{dx}{dt}
   + \epsilon  {\left[{\left( \frac{\p{\eta}}{\p{x}}
   - \frac{\p{\tau}}{\p t} \right)}{ \frac{dx}{dt}} -  {\frac{\p\tau}{\p x}}
   {\left(\frac{\p x}{\p t}\right)}^2 \right]}  
     + O({\epsilon}^2)  \label{eq:dxdt2} .
\end{eqnarray}
A similar procedure is followed to express
$\frac{{d}^2 {\tilde x}}{{d{\tilde t}}^2} $
likewise. By substituting equations (\ref{eq:zwei} ) - (\ref{eq:dxdt2} ) 
for a given explicit expression
for $ {{\omega}_a} $ and remembering that  
$ {\frac{d^2 {{x}_a}}{dt^2}  - {\omega}_a}  $
is zero, we obtain the desired partial differential equation whose solution 
would determine $ {\tau{(t,x)}} $ and $ {\eta{(t,x)}} $. In our case, 
of course,
we have to find $ {\tau{(t, x_1, x_2, x_3 )}} $ and 
$ {{\eta}_a}{(t, x_1, x_2, x_3 )} $'s.
 
To relate these to the generators of the infinitesimal transformations
we write
\begin{eqnarray}
  {\tilde t} {(t, x_i ; \epsilon )} =  t +
  \epsilon {\tau (t , x_i )} +  \cdots 
  = t + \epsilon{ \bf{X}} t + \cdots  \\
  {{{\tilde x}_a} {({{t}{,}{x_i} {;}{\epsilon}}) }} = {{x_a}  +{\epsilon}
  {{\eta} {({{t}{,}{x_i}} )}} + {\cdots}}
   = {{x_a} + {\epsilon} {\bf{X}}{ x_a } + {\cdots}}   
\end{eqnarray}
where the functions $\tau$ and ${\eta}_a $ are components of tangent vectors 
at the
points
${\tilde t }$ and ${{\tilde x}_a} $  defined by
\begin{eqnarray}
   {\tau (t,x_i ) } &=& {{ \frac{\p{\tilde t}}{\p{\epsilon}}}
   {{\mid}_{\epsilon = 0}}} ,\\ 
   {{{\eta}_a}(t,x_i )} &=& {{\frac{\p{\tilde{x_a}}}{\p{\epsilon}}}
    {{\mid}_{\epsilon =0}}}
\end{eqnarray}
and the operator $\bf{X}$ is given by
\begin{eqnarray}
   {\bf{X}} &=& {{\tau (t,x_i )} {\frac{\p}{\p{t}}}  +
   {{\eta}_a (t,x_i )}} {\frac{\p}  {\p{x_a}}}.
\end{eqnarray}
where repeated indices are summed.
Following Stephani \cite{stephani}, we will find the infinitesimal 
generators of the symmetry under which the system of
differential equations do not change.  The 
symmetry is generated by $\bf{X} $ and its extension
\begin{eqnarray}
   {\dot{\bf{X}}} &=& \tau {\frac{\p}{\p{ t}}} + 
   {{\eta}_a} {\frac{\p}{{\p}{x_a}}} + 
   {\dot{\eta}}_a {\frac{\p}{{\p{\dot{x}}_a}}}  
\end{eqnarray}
and the symmetry condition under transformations
represented by equation (\ref{eq:zwei}) determines  ${\dot{\eta}_a} $.
In the expanded form the 
symmetry condition becomes
\begin{eqnarray}
    {\eta}_b {{\omega}_a}_{,b} + {( {{\eta}_b}_{,t}
   + {\dot{x}}_c {{\eta}_b}_{,c} - {\dot{x}}_b  {{\tau}_{,t}}
   -  {\dot{x}}_b {\dot{x}}_c  {{\tau}_{,c}} )}
   { \frac{\p{{\omega}_a}}{\p{\dot{x}}_b}}  
   +  {\tau} {{{\omega}_a}_{,t}} + 2 {{\omega}_a} {( {{\tau}_{,t} + 
   {\dot{x}}_b}{{\tau}_{,b}} )} \nonumber \\
   + {\omega_b}{({\dot{x}}_a {{\tau}_{,b}} - {{\eta}_a}_{,b}) } 
   + {\dot{x}}_a  {\dot{x}}_b  {\dot{x}}_c  {\tau}_{,bc} + 
   {\dot{x}}_a {{\tau}_{,tt}}  
   + 2 {\dot{x}}_a  {\dot{x}}_c {{\tau}_{,tc}}   \nonumber \\ 
    -  {\dot{x}}_c  {\dot{x}}_b  {{\eta}_a}_{,bc}
   - 2 {\dot{x}}_b {{\eta}_a}_{,tb}
   - {{{\eta}_a}_{,tt}}  = 0  \label{eq:condition}
\end{eqnarray}
where $ f_{,t} = {\frac{{\p}f}{\p{t}}} $ and $ f_{,c} 
  = {\frac{{\p}f}{\p{x_c}}} $.
By herding together coefficients of the terms that are
 cubic, quartic, and linear 
in $ {\dot{x_a}} $ ,
and the ones  independent of $ \dot{x_a} $ separately, and equating each of 
these
to zero we obtain an over determined set of partial differential equations
and solve for $ {\tau} $ and $ {{\eta}_{a}} $.

\section{ Symmetry generators, classical particle }

The solutions of the symmetry conditions provide us the generators of the 
group.
In this section, we explicitly obtain the generators for the cases 
mentioned in the introduction. 

For a charged particle moving in the absence of 
electromagnetic field the equation
of motion is given by,
\begin{eqnarray}
m {\ddot{x}}_k  = 0 \label{eq:free}
\end{eqnarray}
and it is well known that the symmetry condition, which is
 equation(\ref{eq:condition}),
gives rise to the eight parameter symmetry generator of the general projective
transformation  as given by
\begin{eqnarray}
  {\bf X } = [ {a_1} + {a_2} t + {a_3} {x_a} + {a_4} t {x_a} + {a_5} {t^2} ]
   {\frac{\p}{\p t}} \nonumber \\  
   + [ {a_6} + {a_7} t + {a_8} {x_a} + {a_5}t {x_a} + {a_4} {{(x_a)}^{2}}]
   {\frac{\p}{\p{x_a}}}  \label{eq:free1} 
\end{eqnarray}
for each of the equations(\ref{eq:free}).
We mention this result, so that the generators can be compared with the 
results we would obtain later for our examples.

We first give the complete analysis of a simpler case, which is the case(i)
as mentiones in the introduction. \\

{\bf {Case (i):}}\\
\\
For the motion of a charged particle in the presence
of a magnetic field proportional to the coordinate vectors, the equations 
of motion are 
\begin{eqnarray}
   {\ddot{x}}_k  &=& {\beta}{\varepsilon}_{kbc}{\dot{x}}_b{x_c} = {\omega}_k 
 \label{eq:mag1}
\end{eqnarray}
where the above expression in the middle  is the Lorentz
 force acting on a charged particle.
The magnetic field for this case is proportional to $ x_c $.

Substituting
\begin{eqnarray}
   {{\omega}_k} &=& {\beta}  {{{{\varepsilon}}_{kbc}}{\dot x_b}{x_c}}
\end{eqnarray}
into equation (\ref{eq:condition}) we obtain, in general, coupled partial 
differential
equations for $ \tau $ and $ \eta $ by equating to zero the terms
corresponding to various powers of $ {\dot x_l} $.

Consideration of the term with $ {\dot{x}}_a {\dot{x}}_b {\dot{x}}_c $
in Eq.(\ref{eq:condition})
 tells us
\begin{eqnarray}
   {{\tau}_{,bc}} &=& 0   .
\end{eqnarray}
Hence we may have
\begin{eqnarray}
    {\tau} &=& {{{A_l}(t)} {x_l} + B(t) + C } .
\end{eqnarray}
The terms quadratic in $ {\dot{x}} $ give
\begin{eqnarray}
    {-}{\beta}{\dot{x}}_b {\dot{x}}_c  {x_m}{{\tau}_{,c}}{{\varepsilon}_{abm}}
    + 2{\beta} {\dot{x}}_l {\dot{x}}_b {q_m}{{\tau}_{,b}}{{\varepsilon}_{alm}}
              \nonumber \\
    + {\beta} {\dot{x}}_r {\dot{x}}_a  {x_s}{{\tau}_{,b}}{{\varepsilon}_{brs}}
    + 2 {\dot{x}}_a  {\dot{x}}_c  {{\tau}_{,tc}}  -  {\dot{x}}_c
    {\dot{x}}_b {{{\eta}_a}_{,bc}} 
    &=& 0
\end{eqnarray}
This shows that $ \tau $ has to be independent of $ x_l $. Hence
\begin{eqnarray}
   {\tau} &=& B{(t)} + C  ,
\end{eqnarray}
and $ \eta $ may have the form
\begin{eqnarray}
   {{\eta}_a} = { D{(t)} {x_a} + {E_l}{(t)} {{\varepsilon}_{lam}} {x_m}
   +  {F{(t)}} + G }  .
\end{eqnarray}
The terms linear in ${\dot x}_l$ provide
\begin{eqnarray}
  {\beta} {{\dot{x}}_l} {{\varepsilon}_{alb}} {{\eta}_{b}}
   + {\beta} {\dot{x}}_c {x_m} {{\varepsilon}_{abm}} {{{\eta}_b}_{,c}}
   - {\beta} {\dot{x}}_b {x_m}{{\varepsilon}_{abm}} {{\tau}_{,t}}  
   + 2{\beta}{\dot{x}}_l  {x_m} {{\varepsilon}_{alm}}  {{\tau}_{,t}} \nonumber \\
   - {\beta} {\dot{x}}_r {x_s} {{\varepsilon}_{brs}}
   {{{\eta}_{a}}_{,b}} 
   - 2 {\dot{x}}_b  {{{\eta}_{a}}_{,tb}}  +  {\dot{x}}_a {{\tau}_{,tt}}   
    = 0    .
\end{eqnarray}

This demands
\begin{eqnarray}
   {B{(t)}} &=& { t H  + C} ,
\end{eqnarray}
and also ${\eta}_a$ has to be independent of $t$, giving
\begin{eqnarray}
   {{\eta}_a} &=& {-} {H} {x_a} + {E_l}{{\varepsilon}_{lam}} {x_m}  .
\end{eqnarray}
Thus we obtain five generators
\begin{eqnarray}
   {{\bf{X}}_a} = {{\varepsilon}_{akb}} {x_b} {\frac{\p}{\p{x_k}}},\qquad
   {{\bf{X}}_4}  = {\frac{\p}{\p{t}}},   \qquad
   {{\bf{X}}_5}  = {t}{\frac{\p}{\p{t}}} - {x_a} {\frac{\p}{\p{x_a}}} 
\end{eqnarray}
and their Lie algebra
\aq
 \left[ {{\bf{X}}_a}, {{\bf{X}}_b} \right] = {{\varepsilon}_{abc}}{{\bf{X}}_c},
  \quad 
 \left[ {{\bf{X}}_a}, {{\bf{X}}_4} \right] = 0, \quad
   \left[ {{\bf{X}}_a}, {{\bf{X}}_5} \right] = 0, \quad
   \left[ {{\bf{X}}_4}, {{\bf{X}}_5} \right] =   {{\bf{X}}_4}.\label{eq:G2IIa} 
\qa
A comparison with the results of similar analysis for the Kepler problem
\cite{stephani} shows that the first four generators are identical, the first
three corresponding to the generators of the three dimensional rotation group
and $ {{\bf{X}}_4} $ is the generator for time translation. However in this 
case
the law corresponding to Kepler's third law goes instead like
\begin{eqnarray}
\tilde{t} \tilde{r} &=& tr = constant.
\end{eqnarray}

The Lie algebra represented by equation (\ref{eq:G2IIa}) corresponds, 
in the notation
of Stephani \cite{stephani}, to the group $ SO(3) \times {{G_2}IIa}$, 
 where  the  $ {{G_2}IIa}$ is a group with the two generators
\aq
{{\bf{X}}_4}  = {\frac{\p}{\p{t}}}, \qquad
 {{\bf{X}}_5}  = {t}{\frac{\p}{\p{t}}} - {x_a} {\frac{\p}{\p{x_a}}}.
\qa

We can find their extension from the formula
\begin{eqnarray}
   {{\dot{\eta}_{a}}} = {\frac{d{{\eta}_a}}{dt}} 
   - {\dot{x}}_a {\frac{d{\tau}}{dt}}
\end{eqnarray}
and obtain, denoting the extensions by $ {\dot{{\bf{X}}}} $,
\begin{eqnarray}
   {\dot{\bf X}}_a  &=&  {{\varepsilon}_{akb}} {( {x_b}{\frac{\p}{\p{x_k}}}
   +  {\dot{x}}_b {\frac{\p}{\p{\dot{x}}_k}} )},  \\
   {\dot{\bf{X}}}_4  &=&  {\frac{\p}{\p{t}}},  \\
   {\dot{\bf{X}}}_5  &=&  { t} {\frac{\p}{\p{t}}} - {x_a}{\frac{\p}{\p{x_a}}}
   - 2 {\dot{x}}_a  {\frac{\p}{\p{\dot{x}}_a}}, \quad scaling. 
\end{eqnarray}

 Henceforth we scale $\beta$, which is a function of the coupling
 constant, 
mass, etc. to one.

Through an  analysis that is similar to the above procedure, we  find the 
 the generators of the symmetry groups for the following examples.
\\

{\bf{Case (ii):}}\\
\\
If the particle is further constrained to move on the surface of a sphere of
unit radius, the equation of motion becomes
\begin{eqnarray}
   {\ddot{x}}_a &=& {{\varepsilon}_{abc}} {{{\dot{x}}_b}} x_c
   - {x_a}{\dot{x}}_k{\dot{x}}_k  \label{eq:wzw}.
\end{eqnarray}
This is equivalent to the case of a particle moving in the presence of
a magnetic monopole centered at the origin of the sphere. Witten has 
generalized this idea to arbitrary dimensions for field theoretic 
considerations. In the above and henceforth we have scaled $\beta$ to
one. 
As has been pointed out by Witten \cite{witten2}, one faces trouble 
in attempting to derive
these equations of motion by using the usual procedure of variation of a
Lagrangian since no obvious term can be included in the Lagrangian whose
variation would give the equation of motion (\ref{eq:mag1}). Hence it
would be more appropriate here to consider the group analysis of the 
equations of motion directly to obtain all the Lie point
symmetries. Usually
the Noether symmetries are a subclass of these. However,
the present analysis cannot give any of the non-Lie symmetries.

With $ {\omega}_a $ being equal to the right hand side of 
equation (\ref{eq:wzw}),
the group analysis shows that there is only one trivial 
time translation besides the generators of the angular momentum
 for this problem,
\begin{eqnarray}
  {{\bf{X}}_a} = {{\varepsilon}_{akb}} {x_b} {\frac{\p}{\p{x_k}}},\qquad
{\bf{X_4}} &=& {\frac{\p}{\p{t}}}  .
\end{eqnarray}

Same is the case if we ignore the term
containing ${\varepsilon}_{abc}$in equation (\ref{eq:wzw}).
So  also for a magnetic field  $ {\bf B}  = -b {x_1} {{\bf{e}}_1}
+ ( B_0 + b {x_3}){ {\bf{e}}_3}  $, which
is an idealized version of the Stern-Gerlach magnetic field \cite{bohm}. 
\\

{\bf{Case (iii):}} \\
\\
 If a potential like  $ \frac{1}{r^2} $ is only present we find
 the generators with extensions to be
\aq
{\dot{\bf {X}}_a } =  {{\varepsilon}_{abc}} \left( {x_c}\frac{\p}{\p x_b}
   + {{\dot x}_c {\frac{\p}{\p {\dot x}_b}}} \right), 
  \quad    space \ rotations,  \nonumber \\
 \dot{{\bf {X}}_4 } =  { \frac{\p}{\p t}}, \quad time \ translation,  \nonumber \\
  {\dot{\bf {X}}_5 }  = 2 t  \frac{\p}{\p t}  + {x_a}\frac{\p}{\p x_a} -
  {\frac{1}{2}}  {{\dot x}_a}
 {\frac{\p}{\p {\dot x}_a}},     \nonumber \\ 
   \  Kepler \ like \ scaling \ law
 \ {\frac{t}{r^2}} = constant, \nonumber  \\
  {\dot{\bf {X}}_6 }  =  t^2  \frac{\p}{\p t}  + t {x_a}\frac{\p}{\p x_a} +
 {{ x}_a}
  {\frac{\p}{\p {\dot x}_a}}  -  {{\dot x}_a}
  {\frac{\p}{\p {\dot x}_a}}       
  \label{eq:mono}
\qa
The vector fields have the commutation relations
\aq
  \left[{\bf X }_{a}, {\bf X }_{b} \right] =
   {{\varepsilon}_{abc}} {\bf X }_{b},  \nonumber \\ 
   \left[{\bf X }_{a}, {{\bf X }_4} \right] = 
 \left[{\bf X }_{a}, {{\bf X }_5} \right] = 
 \left[{\bf X }_{a}, {{\bf X }_6} \right] =  0 \nonumber  \\
 \left[{\bf X }_{4}, {\bf X }_{5} \right] = 2 {\bf X }_{4},\
     \left[  {\bf X }_{4}, {\bf X }_{6} \right] =  {\bf X }_{5},\
   \left[  {\bf X }_{5}, {{\bf X }_{6}} \right] = 2 {{\bf X }_{6}}
\qa

The classical Kepler problem with $ \frac{1}{r} $ potential
has a different scaling law of $ \frac{t^2}{r^3} $ and also 
does not possess the symmetry corresponding to generator
${\bf X_6}$. However, it possesses a Runge-Lenz vector.
Stephani has given a general method to obtain such conserved
vectors in the Lagrangian formulation. In the quantum mechanical case,
if the eigenvalues are taken instead the Hamiltonian operator,
an enhanced symmetry with closed Lie algebra occurs 
for  $ \frac{1}{r} $ potential.
For  $ \frac{1}{r^2} $ potential we could not find a classical Runge-Lenz
vector by Stephani's method. This appears to be related  to
orbits being not closed in such a potential\cite{gold}. However, as 
has already been noted,
in this case new vector fields result leading to the  extra symmetries. 
\\

{\bf{Case (iv):}} \\
\\
Jackiw has considered the symmetries of equation of motion, Lagrangian,
and Hamiltonian for a charged particle in the field of a magnetic 
monopole \cite{{jackiw},{jackiw2}}. He had discovered an extra
  $SO(2,1)$ hidden symmetry by scaling and physical considerations.
Leonhardt and Piwnicki have explored the
theoretical possibility of obtaining the  field of
quantized monopoles when a classical dielectric moves in a charged
capacitor \cite{leonhardt}. 
Since the magnetic field due to a magnetic monopole is $ {B_a} = {{x_a}\over 
{r^3}}  $, the equation of motion is
\aq
{{\ddot{x}}_a } = {{\varepsilon}_{abc}}  {\frac{{{\dot{x}}_b} {x_c}}
  {r^3}}  = {\omega_a} \label{eq:monopl}
\qa
We have taken the coupling constant, mass etc. to be unity.
Lie point symmetries of these equations were obtained in\cite{{Mor},{Haas}}.
We get  the same  generators 
with extension as in the case (iii) for the  $ \frac{1}{r^2} $  potential.
Zwanzinger had considered the motion of a charged particle in the presence of 
monopole along with a  $ \frac{1}{r^2} $  potential\cite{Zwa}. Hence 
 the equation of motion 
\aq
{{\ddot{x}}_a } = {{\varepsilon}_{abc}}  {\frac{{{\dot{x}}_b} {x_c}}
  {r^3}}  + {\frac{ {\mu}^2 {x_a}  }{mr^4}}  \label{eq:zwa}
\qa
also possesses the same above symmetry. The last terms of the equation
reminds of an electric dipole potential at
large distances in the  one dimensional case. 
\\

{\bf{Case(v):}}\\
\\
We have obtained for the case of the field due to both 
a monopole and a charge, i.e. a dyon, only the 
 first four generators  of 
equation (\ref{eq:mono}) for case(iii). 
\\

{\bf{Case (vi):}} \\
\\
But for a velocity dependent potential with equations of motions of the form
\begin{eqnarray}
   {\ddot{x}}_a &=& {\dot{x}}_a {x_k}{x_k} \label{eq:vel} 
\end{eqnarray}
we again find five symmetry generators, the first four being the same
as ${\dot{\bf{X}}}_a $ and $ {\dot{\bf{X}}}_4 $ while
\begin{eqnarray}
   {\hat{\bf X}}_5 &=& 2 t {\frac{\p}{\p{t}}}
   - {x_a}{\frac{\p}{\p{x_a}}}  ,
\end{eqnarray}
and with its extension,
\begin{eqnarray}
   {\dot{\hat{\bf X}}}_5 &=& 2 t {\frac{\p}{\p{t}}} - {x_a}{\frac{\p}{\p{x_a}}}
   - 3 {\dot{x}}_a  {\frac{\p}{\p{\dot{x}}_a}}. \label{eq:dotX5}
\end{eqnarray}
The length and time scale in this case as
\begin{eqnarray}
   {\tilde t}{{\tilde r}^2} &=& t{r^2} .
\end{eqnarray}

{\bf {Case (vii):}} \\
\\
For a charged particle moving in a model magnetic field of the form
\begin{eqnarray}
  {\bf B }  
  = \{ B_x = 0 , B_y = 0 , B_z = {-} {\frac{\cal B} {x^2}} \}  \label{eq:b1}
\end{eqnarray}
 the equations of motion are
\aq
  {{\ddot{x}}_1} = - {\frac{{{\dot x}_2}{\cal B}}{{x_1}^2}}, \qquad
  {{\ddot{x}}_2} =  {\frac{{{\dot x}_1}{\cal B}}{{x_1}^2}}, \qquad
  {{\ddot{x}}_3} = 0.
\qa
 Here $ \cal B $ is a constant. This magnetic field may be obtained 
from a current density $ {\bf J}
= ( 0, {\frac {\cal B}{x^3}}, 0 ) $,
which is singular. It is interesting to note, however, that
 the Schr\"odinger equation can be exactly
solved and corresponding energy levels be obtained in the manner of
Landau \cite{mah03}.
The condition (\ref{eq:condition}) 
for $a = 1$, and $2$ gives
\begin{eqnarray}
 \tau = \lambda , \qquad
 \eta_1 = 0 ,  \qquad  \eta_2 = \sigma ,
\end{eqnarray}
and for $a = 3$ we obtain
\begin{eqnarray}
\eta_3 =  \rho + x_3
\end{eqnarray}
where $ \lambda $, $\sigma$, and $\rho $ are constants.
This gives rise to the vector fields
\begin{eqnarray}
   {\bf X }_{\tau} = \lambda {\frac{\p}{\p t}} ,\quad
   {\bf X }_{{\eta}_2}  = \sigma  {\frac{\p}{\p {x_2}}} ,\quad 
    {\bf X }_{3}  = \rho  {\frac{\p}{\p {x_3}}}, \quad
    {\bf X }_{{\eta}_3} = x_3  {\frac{\p}{\p {x_3}}},
\end{eqnarray}
that forms a solvable Lie algebra. The commutation relations are given 
by
\aq
  \left[{\bf X }_{\tau}, {\bf X }_{{\eta}_2} \right] = 0,  \quad
   \left[{\bf X }_{\tau}, {{\bf X }_3} \right] = 0,  \quad
 \left[{\bf X }_{\tau}, {\bf X }_{{\eta}_3} \right] = 0,   \nonumber  \\
     \left[  {\bf X }_{{\eta}_2}, {\bf X }_{3} \right] = 0, \quad
   \left[  {\bf X }_{{\eta}_2}, {{\bf X }_{{\eta}_3}} \right] = 0 ,\quad
   \left[  {\bf X }_{3}, {\bf X }_{{\eta}_3} \right] =  {\bf X }_{3}.
\qa
and correspond to direct products of two abelian groups and  and the 
group ${G_2} IIb $ which has the generators
\aq
 {\bf X }_{3}  = \rho  {\frac{\p}{\p {x_3}}} \nonumber \\
  {\bf X }_{{\eta}_3} = x_3  {\frac{\p}{\p {x_3}}}.
\qa

We also analyse the Landau problem with a 
constant magnetic field in the
$x_3$ direction. The equation of motion is given by,
\begin{eqnarray}
   {\ddot{x}}_k  &=& {\beta}{\varepsilon}_{kbc}{\dot{x}}_b B_c  = {\omega}_k 
 \label{eq:magL}
\end{eqnarray}
where ${\bf B} = (0, 0, B) $, with $ B$ a constant. 
The vector field obtained is
\aq
 {\bf X }_{1}  =   {\frac{\p}{\p {x_1}}}, \qquad 
  {\bf X }_{2}  =   {\frac{\p}{\p {x_2}}}, \qquad 
{\bf X }_{3}  = x_{1}  {\frac{\p}{\p {x_2}}} -  x_{2}  {\frac{\p}{\p {x_1}}},
 \qquad    {\bf X }_{4}  =   {\frac{\p}{\p {t}}}.
\qa
with commutation relations
\aq
 \left[ {{\bf{X}}_1}, {{\bf{X}}_2} \right] = 0,
  \quad 
 \left[ {{\bf{X}}_1}, {{\bf{X}}_3} \right] =  {{\bf{X}}_2}, \quad
   \left[ {{\bf{X}}_2}, {{\bf{X}}_3} \right] =  -{{\bf{X}}_1}, \quad
   \left[ {{\bf{X}}_a}, {{\bf{X}}_4} \right] =  0.\label{eq:L} 
\qa
 Thus we have found the generators that specifies the corresponding 
 symmetry groups for all our examples considered in the introduction.

\section{ Symmetry generators, quantum particle }

The simplest one dimensional version  of the equation
\aq
{{\ddot{x}}_a } =  {\frac{{ \mu}^2  { x_a}}{mr^4}}  \label{eq:fub}
\qa 
possesses remarkable symmetries which were exploited by de Alfaro,
 Fubini, and Furlan to construct conformal quantum mechanics. Here $x$ is 
considered as a field in
zero space and one time dimension.
The quantum mechanical equation for the wave function $u$
 becomes, in our notation, 
\aq
  {\left( - {\frac{d^2}{{dx}^2}} + {\frac{g}{x^2}} +
 {\frac {x^2}{a^2}} \right)} u  =  {\frac{4r}{a}} u 
\qa
where $   {\frac{{ \mu}^2 }{m}} $ is replaced by $g$. Here $a$ is a
constant which plays a fundamental role in the theory and $r$ is related
to appropriate raising and lowering operators. This equation
 can be expressed in terms of  differential operator realization
of $su(1,1)$ algebra\cite{wybourne} and was studied in detail 
 in\cite{deA}.

There has been earlier works, where it has been shown that the
 dynamics of a scalar particle approaching the event horizon of 
a blackhole is governed by an Hamiltonian with an inverse 
square potential\cite{{berg},{Cl},{Gib},{Wt},{t1},{t2},{Sen}}. The scalar
 field can be 
used as a probe to study the geometry in the vicinity of the horizon
 and its dynamics is expected to provide clues to the inherent 
symmetry properties of the system.
The Hamiltonian of conformal quantum mechanics fits into this. This 
Hamiltonian also arise as a limiting case of the brickwall model 
describing the low energy quantum dynamics of a field in the background 
of a massive Schwarzschild blackhole of mass $M$ \cite{{t1},{t2}}. By 
factorizing such a Hamiltonian,
Birmingham, Gupta and Sen have found the Virasoro symmetry of the system
and have studied the representation of the algebra as well as the scaling 
properties of the time independent modes\cite{Sen}.
They had obtained the full Virasoro algebra by the requirement of unitarity
of the representation. The Hamiltonian operator is in the enveloping algebra. 
However, here we aim to find the underlying Lie point 
symmetry 
of the equation of motion of the scalar particle, viewed as a 
differential equation. 
 Of course, mathematically 
any two linear homogeneous
ordinary differential equations can be transformed to the form,where a 
prime denotes a differentiation with respect to
$x$,
\aq
   u''  = 0.
\qa
 This equation has the eight dimensional symmetry of projective 
transformations.  But the two equations could be different from the physics 
point of view having different eigenvalues and eigenfunctions. Hence
 we would like to see explicitly what are the Lie point symmetries of the
particular equation.

For the equation
\begin{eqnarray}
  u'' = \omega (x,u,u') \label{eq:our}
\end{eqnarray}
where
\begin{eqnarray}
{\omega} (x,u,u') = - { ( {C\over{x^2}} + {D\over{x}}
  + \hat{E})} u(x).\label{eq:CDE}
\end{eqnarray}
the symmetry generators are obtained from the conditions given by the
equation  (\ref{eq:condition}) which reduces in the one dimensional case to
\begin{eqnarray}
\omega ( {\eta}_{,u} - 2 {\tau}_{,x} -3 u' {\tau}_{,u} )
 - {\omega}_{,x} \tau - {\omega}_{,u} {\eta}
- {\omega}_{,{u'}} {[ {\eta}_{,x} + u' ( {\eta}_{,u} -{\tau}_{,x} )
 - {u'}^2 {\tau}_{,u} ]}     \nonumber     \\
+ {\eta}_{,xx} + u' ( 2 {\eta}_{,xu} - {\tau}_{,xx} )
+ {u'}^2 ( {\eta}_{,uu} - 2 {\tau}_{,xu} ) - {u'}^3 {\tau}_{,uu}
= 0  \label{eq:oned}
\end{eqnarray}
For the case 
$C = - g $, $ D = \hat{E}  = 0 $,
equating to zero the coefficients of ${u'}^3 $ and ${u'}^2 $ in
(\ref{eq:oned})
we get
\begin{eqnarray}
{\tau}_{,uu} =0 , \qquad {\eta}_{,uu} = 2 {\tau}_{,xu}
\end{eqnarray}
which are satisfied for
\begin{eqnarray}
{\tau} = u \alpha (x) + \beta (x) , \qquad  \eta  = u^2  {\alpha}' (x) +
 u {\gamma}(x) + \delta (x)  .
\end{eqnarray}
Using these and equating to zero the coefficient of $u'$ 
 and then considering the 
 the terms not involving  $u'$, we find that
an interesting symmetry exists only when the coupling constant $g$ 
is equal to 2. For this case
 we obtain 
\aq
 {\tau} = {\frac{1}{x}} Au + Fx \nonumber \\
{\eta} =  -  {\frac{1}{x^2}} Au^2 + B u + {\delta}(x) \label{eq:sen}
\qa
where $ A, F$, and $B$ are constants and ${\delta}(x)$ satisfies 
the same equation as $u$ does. The vector fields are
\begin{eqnarray}
{{\bf X}_1} = x {\frac{d}{dx}}, \qquad {\bf X_2} = u {\frac{d}{du}},  \qquad
{{\bf X}_3} =  {\frac{1}{x}}u {\frac{d}{dx}}
  -  {\frac{1}{x^2}}u^2 {\frac{d}{du}} \label{eq:sn}
\end{eqnarray}
with commutation relations
\aq
[{{\bf  X}_1 } ,{ {\bf X}_2 } ] = 0, \qquad 
 [{{\bf  X}_1 } , {{\bf X}_3 } ] =  -2  {{\bf X}_3 }, \qquad  
[{{\bf  X}_2 } , {{\bf X}_3 } ] =  {{\bf X}_3 }
\qa
For the  case,  considered in \cite{Sen2}, 
  the relevant equation is
\begin{eqnarray}
 {\frac{d^2 u}{{dx}^2}}  + {\frac{1}{x^2}} 
 {\left[{\frac{1}{4} + {R^2}{E^2}}  \right]} u = 0 \label{S0}
\end{eqnarray}
where $E$ is a generic eigenvalue and $R = 2M$.
Hence $g$ corresponds to ${\left[{\frac{1}{4} + {R^2}{E^2}}\right]} $
 in this case.
However, our result shows that  only when $E$ is imaginary
with  ${\left[{\frac{1}{4} + {R^2}{E^2}}\right]} = -2 $, the symmetry
will show up. Further, the $ {\frac{1}{x}}$
and $ {\frac{1}{x^2}}$ factors in ${{\bf X}_3}$ makes it ill defined as
$x \rightarrow 0 $ similar to the $L_{-n}$ operators of conformal field theory 
or the $ P_m $ operators considered by Birmingham, Gupta, and Sen\cite{Sen}.

\section{Complete Krause symmetry }

Krause has introduced the concept of the complete symmetry group of a system by
specifying two extra properties in the definition of a Lie symmetry group.
This requires the manifold of solutions to be a homogeneous space on which 
the group action takes place and the group is specific to the system with no
other system admitting it.

Besides the Lie point symmetries and the contact symmetries, new types of 
symmetries are to be included in order to obtain the complete symmetry 
group\cite{krause}. For an N dimensional system, the generators of the 
new symmetry was defined to be
\begin{eqnarray}
  Y = \left[ \int {\xi}(t, x_1, x_2, \cdots , x_N) dt \right] {\p}_t 
  + \sum_{k=1}^{N} {\eta}_k (t, x_1, x_2, \cdots , x_N) 
   {\p}_{x_k}    \label{eq:Y}
\end{eqnarray}
which is different from the generators of a Lie point transformation
because of the appearance of the integral of $\xi$. This makes it 
a nonlocal operator. 

Nucci has developed a method based on the reduction of order to derive 
all Lie symmetries\cite{nucci}. Later Nucci and Leach  have found the 
existence of 
more nonlocal symmetries. These symmetries become local on reduction
of order. In this technique, for an autonomous system, one of
the unknown function is taken as the new independent variable and the 
system  is written in the modified form. Then the standard Lie group 
analysis of this transformed system yields the extra symmetries leading 
to a complete attainment 
of Krause symmetries. The method can be extended to include 
non-autonomous systems\cite{nuccikepler2}. Using a different
technique, it has also again being found that the three dimensional 
Kepler problem is completely specified by six symmetries\cite{nuccimar}.

Besides the original Kepler problem, this method has been used to 
analyze many  problems of physics, space science, 
meteorology etc. which include the Kepler problem with a drag, 
motion in an angle dependent force\cite{{nuccikepler2},{andrio}}, Lorenz 
equation\cite{nuccilorenz}, Euler-Poinsot system and Kowalevsky 
top\cite{{nuccijacobi},{nuccimar}}, as well as
 relativistically spherically symmetric 
systems\cite{cot}. Nucci's  interactive code for determination of Lie 
symmetries  has been used  
to arrive at the above results\cite{nuccireduce}.

We follow the method and notation of references\cite{{nucci},{nuccikepler2}} 
to determine the
Krause symmetries for our examples.
In our case the system of differential equations are given by
\begin{eqnarray}
  {\ddot{x}}_k = F_k ( x_1, x_2, x_3, {\dot{x}}_1,  {\dot{x}}_2,  {\dot{x}}_3 )
        \label{eq:Fk}
\end{eqnarray}
where $ k = 1,2,$ and $3$ and the $ {\omega}_k $ of equation(\ref{eq:nfree})
equals $F_k$.
By standard techniques\cite{{ovs},{ibragimov}}  one obtains the generators
of the Lie point group for this system and a generator is written
in the form
\begin{eqnarray}
    X = \tau (t, x_1, x_2, x_3) {{\p}_t +  
    \sum_{k=1}^{3} {\eta}_k (t, x_1, x_2, x_3) {\p}_{x_k} }
\end{eqnarray}
To treat the velocities in the same footing as the coordinates and for 
 reduction of order, equation(\ref{eq:Fk}) is next made into a set of 
six  
ordinary differential equations
\aq
   {\dot{u}}_{k}  = u_{3+k} \label{eq:uk}
\qa
\aq
  {\dot{u}}_{3+k} =  F_k ( u_1, u_2, u_3, u_4, u_5, u_6 )  \label{eq:u3k}
\qa
where $u_k$s are related to $ x_k$s and ${{\dot{x}}_k}$s.
Then one of the dependent variables, $u_i$s, is chosen as the new independent
variable $y$. We take $u_3 = y$. The system (\ref{eq:uk})-(\ref{eq:u3k})
is now converted to a  set of five ordinary differential 
equations  depending on the variable $y$, with
\aq
   {\frac{du_j}{dy}}  = {\frac{u_{3+j}}{u_6}}, \label{eq:dujdy}
\qa
\aq
   {\frac{du_{3+j}}{dy}} =  {\frac { F_j ( u_1, u_2, y, u_4, u_5, u_6 )}{u_6}} 
   \label{eq:du3j},
\qa
\aq
   {\frac{du_6}{dy}}  =  {\frac{ F_3 ( u_1, u_2, y, u_4, u_5, u_6 )}{u_6}},
   \label{eq:du6dy}
\qa
where $j = 1,2 $. Using equation (\ref{eq:dujdy}) we obtain
\aq
   u_{3+j} = u_6 {\frac{du_j}{dy}} \label{eq:dujdy1}.
\qa
This is put back in equations (\ref{eq:du3j}) and (\ref{eq:du6dy}) to give 
the two ordinary second order equations and one first order equation for
the unknowns $u_j = u_j(y)$, and $u_6 = u_6(y)$ 
\aq
{ {{u_j}''} = }  
  {\frac { \left[ { F_j ( u_1, u_2, y, {{u'}_1}, 
  {{u'}_2}, u_6 )}   
   -  F_3 ( u_1, u_2, y, {{u'}_1}, 
  {{u'}_2}, u_6 ) {{u'}_j} \right]  } {{u_6}^2} },   \label{eq:u,,}
\qa
\aq
  {{u'_6}} =  {\frac{1}{{u_6}}}  F_3 ( u_1, u_2, y, {{u'}_1}, 
   {{u'}_2}, u_6 ),  \label{eq:u6,}
\qa
where a prime denotes differentiation with respect to $y$.
For the above system we write a generator for the Lie symmetry group 
as
\aq
   Z = V( y, u_1, u_2, u_6 ) {\p}_{y} + 
   \sum_{j=1}^{2} G_j ( y, u_1, u_2, u_6 ) {{\p}_{u_j} } 
   + G_6 ( y, u_1, u_2, u_6 ) {{\p}_{u_6}}. \label{eq:Z}
\qa
These can be transformed to the old form, $Y$, of the operators of 
equation (\ref{eq:Y}) by replacing $u_j, y, u_6 $ with $x_j, x_3, {\dot{x_3}}$,
respectively, and solving  the following system of equations for
$\xi$ and ${\eta}_k$
\aq
   Y(x_j) \equiv {\eta}_j = G_j,
\qa
\aq
   Y(x_3) \equiv {\eta}_3 = V,
\qa
\aq
   Y^{(1)} \equiv {\frac{d{\eta}_3}{dt}} - {\xi}{\dot{x_3}} = G_6, \label{eq:Y1}
\qa
with $Y^{(1)}$ being the first prolongation of Y.

On application of the above technique to our example 
corresponding to the equation of motion (\ref{eq:mag1}),
\begin{eqnarray}
  {\ddot{x}}_k  &=& {\varepsilon}_{kbc}{\dot{x}}_b{x_c}
  = {\omega}_k  \nonumber \\
\end{eqnarray}
we obtain the following set of equations.
\aq
 { {\Omega}_1} \equiv  {{u_1}''} = 
 {\frac{1}{{u_6}^2}} \left[ ({u_2}' y - u_2 ) - {u_1}'
    ({u_1}' u_2 - {u_2}' u_1 ) \right]  \label{eq:u1}
\qa
\aq
  { {\Omega}_2}  \equiv    {u_2}'' = 
  {\frac{1}{{u_6}^2}} \left[ ({u_1} -  {u_1}' y ) - {u_2}'
  ({u_1}' u_2 - {u_2}' u_1 ) \right]    \label{eq:u2}
\qa
\aq
  { {\Omega}_6}   \equiv    {u_6}' = 
 {1} \left( {u_1}' u_2 - {u_2}' u_1 \right) \label{eq:u6}
\qa
For the determination of $\xi$ and the $\eta$'s 
we note that, first denoting ${\eta}_3 = \zeta $,
\aq
    {\frac {\p {\tilde{u_l}}}{{\p {\tilde y}}} } = {\frac{\p{u_l}}{\p{ y}}}
    + {\epsilon} \left[ {{\eta}_{l,y}} - {{\zeta}_{,y}} {{u_l}'} 
     + \left( {{\eta}_{l,u_i}} - {{u_l}'} {{\zeta}_{,u_i}} \right)
    {{u_i}'} \right]  \label{eq:ubyy}
\qa
and, for example,we find
\aq
   {{{\eta}_6}'} = {{\eta}_{6,y}} + \left( {{\eta}_{6,u_i} } - 
   {{\zeta}_{,{u_{i}}}   {u_6}'}
    \right) {{u_i}'} - {{\zeta}_{,y}} {{u_6}'}
\qa
 with ${{{u_6}'} = {{\Omega}_6}}$.  Denoting the vector-fields 
by $ \bar{\bf X} $, the symmetry condition for our first order 
equation is given by
\aq
  {\bar {\bf X}}{{\Omega}_6} =   
 {{\eta}_{6,y}} + \left( {{\eta}_{6,u_i}   }
   - {{\zeta}_{,u_i}} {\Omega_6} \right)
   {{u_i}'} - {{\zeta}_{y}} {{u_6}'} \label{eq:eta6}
\qa
where
\aq
  {\bar {\bf X}}  = {\zeta}\left({y},{u_i} \right) {\frac{\p}{{\p}{y}}} + 
          {{\eta}_{j}} \left({y}, {u_i} \right)   {\frac{\p}{{\p}{u_j}}} 
\qa
and for the second order equations the symmetry conditions take the form
\aq
  \zeta {{\Omega}_{1,y}} + {\eta}_1  {{\Omega}_{1,u_1}} + {\eta}_2   
   {{\Omega}_{1,u_2}}  + {\eta}_6  {{\Omega}_{1,u_6}}  
\nonumber  \\
   +  \left[ {{\eta}_{1,y}} + \left( {{\eta}_{1,u_i}}  -  {{\zeta}_{,u_i}}
   {{u_1}'} \right) {{u_i}'} - {{\zeta}_{,y}} {{u_1}'} \right]
  {{\Omega}_{1,{{u_1}'}}}   
 \nonumber      \\
   +  \left[ {{\eta}_{2,y}} + \left( {{\eta}_{2,u_i}}  -  {{\zeta}_{,u_i}}
   {{u_2}'} \right) {{u_i}'} - {{\zeta}_{,y}} {{{u'}_2}} \right]
  {{\Omega}_{1,{{{u_2}'}}}}  \nonumber \\
   - \left[ {{\eta}_{1,u_1}} - 2 {{\zeta}_{,y}} - 3 {{\zeta}_{,u_i}} {{u_i}'}
  \right] {{\Omega}_1}   
   - {{\eta}_{1,yy}} - \left( 2 {{\eta}_{1,yu_i}} - {{\zeta}_{,yu_i}} \right)
   {{u_i}'}  
   + {{\zeta}_{,yy}} {{u_1}'} \nonumber \\
    + {{\zeta}_{,u_i}} {{u_1}'} {{\Omega}_i} 
    - {{\eta}_{1,u_j u_i}} {{u_i}'} {{u_j}'} 
    +  {{\zeta}_{,yu_i}}{{u_1}'} {{u_i}'}
    +  {{\zeta}_{,u_i u_j}}{{u_1}'} {{u_i}'} {{u_j}'}
    = 0.   \label{eq:G}
\qa
In the above the appropriate equations  (\ref{eq:u1}),(\ref{eq:u2}),
 and (\ref{eq:u6}) 
are to be substituted.
For equations({\ref{eq:eta6}}) and ({\ref{eq:G}}) to be compatible
we have
\aq
 \zeta = y, \qquad  { {\eta}_1} = {u_1}, \qquad  { {\eta}_2} = {u_2} 
\qa
and we obtain
\aq
  {\xi} = -1
\qa

For our example of equation (\ref{eq:vel})
which is
\aq
   {{\ddot{x}}_a} = {{\dot{x}}_a} {x_k}{x_k} \nonumber \\
\qa
 the equations  for $u_i$'s become
\aq
 { {\Omega}_1} \equiv {{u_{1}}''} = 0, \qquad
   { {\Omega}_2}\equiv  {{u_{2}}''} = 0, \qquad
   { {\Omega}_6}\equiv {{u}_{6}}' =  {{u_1}^2} +  {{u_2}^2} + {y^2}.
\qa
The compatibility of the above three equations forces ${\xi}$ and $\eta$'s
to be
\aq
  {\zeta}  = y,  \qquad
  {{\eta}_1} = {u_1}, \qquad 
  {{\eta}_2} = {u_2}, \qquad 
  {{\eta}_6} = 3{u_6}.
\qa
Thus the  vector field is given by,
\aq
  {\bar {\bf X}} =  {y {\frac{\p}{{\p}{y}}}} + {u_1  {\frac{\p}{{\p}{u_1}}}} + 
         {u_2 {\frac{\p}{{\p}{u_2}}}} +  3{u_6 {\frac{\p}{{\p}{u_6}}} }
\qa
which has a corresponding part in $ {\dot{\bf{X}}}_5 $ of 
equation(\ref{eq:dotX5}). From equation (\ref{eq:Y1}) we obtain $\xi$
to be
\aq
  {\xi}  = -2.
\qa

In the case of a charged particle in the magnetic field of a monopole,
with the equation of motion given by (equation ({\ref{eq:monopl}})),
\aq
{{\ddot{x}}_a } = {{\varepsilon}_{abc}}  {\frac{{{\dot{x}}_b} {x_c}}
  {r^3}}  = {\omega_a} \nonumber
\qa
we get
\aq
 { {\Omega}_1} \equiv  {{u_1}''} =  
  {\frac{1}{u_6}} \left( {u_2}  -
   {u_2}' y -  {{{u_1}'}^{2}}
     u_2 - { u_1}{{ u_1}'}  {u_2}'   \right) 
\qa
\aq
 { {\Omega}_2} \equiv  {{u_2}''} =  
  {\frac{1}{u_6}} \left( 
   {u_1}' y - {u_1}   + {{u_1}'
     u_2 {u_2}'} - { u_1}   \right) 
\qa
\aq
 { {\Omega}_6} \equiv  {{u_6}'} =   {u_1}  {u_2}' -  {u_2}  {u_1}'
\qa
The symmetry conditions are satisfied for
\aq
  {\zeta}  = y,  \qquad
  {{\eta}_1} = {u_1}, \qquad 
  {{\eta}_2} = {u_2}, \qquad 
  {{\eta}_6} = -{u_6}
\qa
and this gives rise to the vector field
\aq
  {\bar {\bf X}} =  {y {\frac{\p}{{\p}{y}}}} + {u_1  {\frac{\p}{{\p}{u_1}}}} + 
         {u_2 {\frac{\p}{{\p}{u_2}}}}   - {u_6 {\frac{\p}{{\p}{u_6}}} }.
\qa
Consequently one obtains,
\aq
  \xi = 2.
\qa

For the case of the last example,i.e, the Landau problem with a constant
magnetic field,  the equations of motion are
\aq
  { {\ddot{x}}_k} = {{\varepsilon}_{klm}}{{\dot{x}}_{l}} {B_m}
\qa
with $ {\bf B} = \left(0, 0, {\frac{1}{x^2}}\right) $. After reduction of 
order the equations are
\aq
  { {\Omega}_1} \equiv  {{u''}_1} 
   = - {\frac{{\cal B}{{u''}_2}}{{u_6}{u_1}}}, \qquad
  { {\Omega}_2} \equiv   {{u''}_2 } 
    =  - {\frac{{\cal B}{{u_1}'}}{{u_6}{{u_1^{2}}}}},    \qquad
   { {\Omega}_6} \equiv  {{u_6}'}    = 0.
\qa

 The analysis results in 
\aq
{\zeta} = y , \qquad
  {{\eta}_1}  = {u_1}, \qquad
 {{\eta}_6}  = - {u_6}.
\qa 
 So we get, also for this case,
\aq
  {\xi} = 2.
\qa
This gives us the vector fields
\aq
   {{\bf {v}}_1} =  {y {\frac{\p}{{\p}{y}}}}, \quad 
   {{\bf {v}}_2} =  {u_1 {\frac{\p}{{\p}{u_1}}}}, \quad 
   {{\bf {v}}_3} = 
    - {u_6}{\frac{\p}{{\p}{u_6}}},  
\qa
So we get back the vector field  $ {\bf X }_{{\eta}_3} $ in ${\bf v}_1$,
but there are now two extra vector fields  ${\bf v}_2$ and ${\bf v}_3$,
which are to be included in the complete symmetry group.

 Thus we have found the generators that specifies the corresponding 
symmetry groups for all our examples as well as the generators of the complete
symmetry of Krause.
  In our  examples, we find that ${\eta}_k $'s do not
depend on $ {\dot{x}}_N $ and $\xi$ is a constant, and hence
$Z$ can be transformed into a generator of a Lie point symmetry group.

\section{Conclusion}

As has already been explicitly mentioned , the equations of 
motion of a free particle admit eight symmetries for 
each of the ${x_a}$s. This
is the maximum number of symmetries for an
ordinary second order differential equation. By including different
$ {x_a} $, $ {\dot{x_a}} $ dependent terms in the equations we
do explicitly see which generators survive as symmetries and we have found
corresponding  complete Lie algebras. We have chosen
some cases motivated by problems from
physics. The original motivation of including Wess-Zumino terms
in the Lagrangian has been to reduce some of its symmetries\cite{witten2},
and here we 
 find that  the equations of motion now support the three dimensional
 rotations and a time translation symmetry instead of the 
six vector fields as for the
 monopole problem without any constraint.
 For the other examples considered here, we get some 
interesting
result in the form of Kepler's scaling law and the full structure
of the inherent symmetry group. This we get without solving
the equations of motion. In the cases where a Lagrangian could be set up, 
those generators operating on the Lagrangian  giving zero include
all the Noether symmetries\cite{stephani}.

It is also expected that related group  analysis may provide useful information
when terms are modified in the Lagrangian, due to quantum corrections,
for example. We have shown explicitly  how many and which generators remain
as symmetries. These symmetries correspond to the transformations of
the solutions. Similar analysis for specific cases of some nonlinear
equations
arising out of linear equations have also been carried out in \cite{mahE}.
We expect that these will ultimately lead to a better
understanding of the spontaneous symmetry breaking, as well.

We have also carried out the analysis using Nucci's method of
reduction of order to find the complete Krause symmetries
in four of our examples. It is found that in our last two examples the 
generators of the complete symmetry group can be transformed into a 
generator of a Lie point symmetry group.

For the case of a scalar particle probing the near horizon structure of a
blackhole, under certain limits the Hamiltonian contains $\frac{1}{r}$
and $\frac{1}{r^2} $ potentials. It is found that there exist a symmetry with
three generators only for specific constant value of the coefficient of the
$\frac{1}{r^2}$ term and in the absence of $\frac{1}{r}$ term. 

The quantum mechanical problem of a charged particle
in the presence of even a constant magnetic field has many interesting 
mathematical structures \cite{plyus} and under certain limits can make 
space coordinates noncommutative \cite{bigatti}.  Klishevich and Plyuschay 
have found a universal algebraic 
structure at the quantum level for the two dimensional case in the
presence of certain magnetic fields\cite{plyus}.  Nonlinear superconformal
symmetry of the fermion-monopole system has been 
extensively studied in\cite{plyus1}. It   would be a 
motivation to seek the existence of  analogous structures for our 
cases.
It would be interesting to study
these quantum aspects as well as the non-Abelian quantum kinematics
in the framework of group theoretic quantization programme of Krause
\cite{krause2} for above nonuniform magnetic fields.\\

\noindent{\bf Acknowledgement}
{ I would like to thank Professor Dieter L\"ust for the warm
hospitality and for providing the academic
facilities at the Arnold Sommerfeld Centre, Ludwig Maximilians
University, Munich, where this work has been done. I am grateful to 
Professor F. Haas for pointing out the reference {\cite{Mor}} to me. }


\end{document}